\documentstyle[a4,12pt,epsf]{article}

\begin{document}

\newcommand{\elf}{ELFE}
\newcommand{\beq}{\begin{equation}}
\newcommand{\eeq}{\end{equation}}
\newcommand{\beqa}{\begin{eqnarray}}
\newcommand{\eeqa}{\end{eqnarray}}
\newcommand{\half}{\frac{1}{2}}
\newcommand{\gsim}{\buildrel > \over {_\sim}}
\newcommand{\lsim}{\buildrel < \over {_\sim}}
\newcommand{\ie}{{\it ie}}
\newcommand{\eg}{{\it eg}}
\newcommand{\cf}{{\it cf}}
\newcommand{\etal}{{\it et al.}}
\newcommand{\gev}{{\rm GeV}}
\newcommand{\jpsi}{J/\psi}
\newcommand{\order}[1]{${\cal O}(#1)$}
\newcommand{\morder}[1]{{\cal O}(#1)}
\newcommand{\eq}[1]{Eq.\ (\ref{#1})}
\newcommand{\ptr}{p_T}
\newcommand{\as}{\alpha_s}
\newcommand{\ket}[1]{\vert{#1}\rangle}
\newcommand{\bra}[1]{\langle{#1}\vert}
\newcommand{\qpair}{q\bar q}
\newcommand{\cpair}{c\bar c}

\newcommand{\PL}[3]{Phys.\ Lett.\ {{\bf#1}} ({#2}) {#3}}
\newcommand{\NP}[3]{Nucl.\ Phys.\ {{\bf#1}} ({#2}) {#3}}
\newcommand{\PR}[3]{Phys.\ Rev.\ {{\bf#1}} ({#2}) {#3}}
\newcommand{\PRL}[3]{Phys.\ Rev.\ Lett.\ {{\bf#1}} ({#2}) {#3}}
\newcommand{\ZP}[3]{Z. Phys.\ {{\bf#1}} ({#2}) {#3}}
\newcommand{\PRe}[3]{Phys.\ Rep.\ {{\bf#1}} ({#2}) {#3}}

\begin{titlepage}
\begin{flushright}
        NORDITA--97/25 P\\
        hep-ph/9703462\\
\end{flushright}

\vskip 2.5cm

\centerline{\Large \bf Physics at ELFE\footnote{Based on talks
given at the meeting on {\em Future Electron Accelerators and Free Electron
Lasers}, Uppsala, April 25-26, 1996 and at the {\em Second ELFE Workshop},
Saint Malo, September 23-27, 1996. Work supported in part by the EU/TMR
contract ERB FMRX-CT96-0008.}}

\vskip 1.5cm

\centerline{\bf Paul Hoyer}
\centerline{\sl Nordita}
\centerline{\sl Blegdamsvej 17, DK--2100 Copenhagen \O, Denmark}

\vskip 2cm

\begin{abstract}
I review some central physics opportunities at the $15\ldots 30$ GeV
continuous beam electron accelerator \elf, proposed to be built in conjunction
with the DESY linear collider. Our present detailed knowledge of
single parton distributions in hadrons and nuclei needs to be supplemented
by measurements of compact valence quark configurations, accessible through
hard exclusive scattering, and of compact multiparton subsystems which
contribute to semi-inclusive processes. Cumulative ($x>1,\ x_F>1$) processes
in nuclei measure short-range correlations between partons belonging to
different nucleons in the same nucleus. The same configurations may
give rise to subthreshold production of light hadrons and charm. The
challenges of understanding high energy charmonium production indicate that
charmonium will be a sensitive probe of color dynamics at ELFE. At low
energies, charmonium forms inside the target nucleus, allowing a
determination of $c\bar c$ bound state interactions in nuclear matter. 
\end{abstract}

\end{titlepage}

\newpage
\renewcommand{\thefootnote}{\arabic{footnote}}
\setcounter{footnote}{0}
\setcounter{page}{1}

\section{Introduction} \label{introduction}

The ELFE@DESY project aims at utilizing a future DESY linear electron
collider~\cite{wiik} to accelerate electrons to $15\ldots 30$ GeV and then use
the HERA electron ring to stretch the collider bunches into an intense
(30 $\mu$A) continuous extracted beam~\cite{frois}. Polarized electrons
will be scattered from both light and heavy fixed targets, with
luminosities in the ${\cal L}= 10^{35} \ldots 10^{38}$ cm$^{-2}$s$^{-1}$
range. In this talk I discuss some of the central physics issues that
can be addressed with this type of accelerator. Since \elf\ experiments are
many years in the future I shall concentrate on questions related to basic
descriptions of the structure of matter at the \order{0.1 {\rm\ fm}} scale in
terms of QCD. These questions will remain of fundamental interest and
require the capabilities of an accelerator like \elf.

\begin{table}[hb]{\centerline{{\rule[-3mm]{0mm}{8mm}}
Table 1. {\bf Features and opportunities of an \elf\ accelerator.}}}\bigskip
\begin{tabular}{p{2cm}llll}{\rule[-3mm]{0mm}{8mm}}
&{\bf Features} & {\bf Opportunities} &\\
\cline{2 - 3} {\rule[-3mm]{0mm}{8mm}}
&High luminosity & Study rare configurations &\\
&$\cal{L}\sim$ $10^{35} \ldots 10^{38}$ cm$^{-2}$s$^{-1}$ &\ of target wave
function &\\
\cline{2 - 3} {\rule[-3mm]{0mm}{8mm}}
&Energy & Perturbative QCD &\\
&$E=15 \ldots 30$ GeV & Resolution of \order{0.1\ {\rm fm}} &\\ 
& & Charm production &\\
\cline{2 - 3} {\rule[-3mm]{0mm}{8mm}}
&High duty factor $\sim 90\%$ & Event reconstruction &\\
\cline{2 - 3} {\rule[-3mm]{0mm}{8mm}}
&Good energy resolution & Exclusive reactions &\\
&$\Delta E/E \simeq 10^{-3}$ & Inclusive reactions at high $x$ &\\
\cline{2 - 3} {\rule[-3mm]{0mm}{8mm}}
&Polarization & Amplitude reconstruction &\\
& & Spin systematics of QCD &\\
\cline{2 - 3} {\rule[-3mm]{0mm}{8mm}}
\end{tabular}
\end{table}

In Table 1 I list the main features of the \elf\ accelerator, and the
opportunities that they provide. Compared to existing electron and muon beams,
the advantages of \elf\ are in luminosity (compared to the muon beams at CERN and
Fermilab), in duty factor (compared to SLAC) and in energy (compared to TJNAF).
Competitive \elf\ experiments will rely on a combination of these strong features.
The HERMES experiment at DESY works in the same energy range but at a lower
luminosity and duty factor compared to ELFE. HERMES will prepare the ground
for \elf\ physics, together with experiments at TJNAF in the U.S., GRAAL in
Grenoble and the lower energy electron facilities ELSA (Bonn) and MAMI (Mainz).

As I shall discuss below, an important part of physics at \elf\ will deal with
exclusive reactions, or with inclusive reactions at large values of Bjorken
$x=Q^2/2m\nu$. The energy range of $15\ldots 30$ GeV is actually optimal for
such studies, as seen from the following argument. The inclusive deep inelastic
cross-section scales (up to logarithmic terms) in the virtuality $Q^2$
and energy $\nu$ of the photon like  
\beq
\frac{d^2\sigma_{DIS}}{dQ^2dx} \propto \frac{1}{Q^4}F(x) \label{siginc}
\eeq
Exclusive processes are still more strongly suppressed at large $Q^2$,
\eg,
\beq
\frac{d\sigma}{dQ^2}(ep\to ep) \propto \frac{F_p^2(Q^2)}{Q^{4}} \propto
\frac{1}{Q^{12}}\ \ \ \ \  (x=1) \label{sigexc}
\eeq
Typically we want to
reach at least $Q^2=\morder{10\ \gev^2}$ to be able to use perturbative QCD
(PQCD) and to have a resolution of \order{0.1\ {\rm fm}}. This implies
$\nu=\morder{5\ \gev}$ at large $x\simeq 1$. At \elf, such energies correspond
to the photon taking a moderate fraction $y=\nu/E_e \simeq 0.15 \ldots 0.3$ of
the electron energy, which is practical for measurements. This may be
contrasted with the situation at HERA, which is equivalent to a fixed target
experiment with an electron energy $E_e\simeq 50000$ GeV. A photon with energy
$\nu=5$ GeV would at HERA correspond to $y \simeq 0.0001$. It is clearly very
difficult to measure the large $x$, moderate $Q^2$ region at HERA, but it is
the natural territory of an accelerator in the \elf\ energy range.

In the following I shall discuss three aspects of physics at \elf\ which
relate to basic issues in QCD:
\begin{itemize} 
\item[$\bullet$] {\em Wave function measurements.} Most of our present
knowledge of hadron and nuclear wave functions stems from
hard inclusive scattering, which measures single parton
distributions. The phenomenology of hard exclusive scattering,
which is sensitive to compact valence quark configurations, is still in its
infancy. Although considerable progress may be expected in this field in
the coming years, the measurements are so demanding that an accelerator with
\elf's capabilities is sorely needed. On the theoretical front, we still do
not have a full understanding of which properties of the wave function are in
principle measurable in hard scattering. It seems plausible that
semi-inclusive processes can be used to  measure configurations where a
subset of partons are in a compact configuration, while the others are summed
over.

\item[$\bullet$] {\em Short range correlations in nuclei.} Scattering which is
kinematically forbidden for free nucleon targets has been experimentally
observed, and includes DIS at $x>1$, hadron production at Feynman $x_F>1$ and
subthreshold production processes. Such scattering requires short range
correlations between partons in more than one nucleon, and thus gives
information about unusual, highly excited nuclear configurations.

\item[$\bullet$] {\em Charm production near threshold\footnote{This was
the main topic of my talk at St. Malo. In this contribution, charm is
discussed in a more detailed form, compared to the other ELFE physics
issues.}.} Production close to threshold requires efficient use of the target
energy and hence favors compact target configurations. Heavy quarks are
created in a restricted region of space-time, where perturbative calculations
are reliable. Both features conspire to make the production of charm near
threshold a sensitive measure of new physics, including unusual target
configurations and higher twist contributions. The \elf\ accelerator will work
in the region of charm threshold $(E_\gamma \simeq 9\ \gev)$ and provide
detailed information about both charmonium and open charm production.
\end{itemize}

The above selection of physics topics is obviously far from complete. I refer
to earlier presentations of \elf\ physics \cite{elphys} as well as to the
review by Brodsky \cite{bro} for further discussions of these and other
aspects of QCD phenomenology. In particular, I shall not cover here the
important and topical area of color transparency, but refer to recent
reviews \cite{ct} and references therein.

\section{Wave function measurements} \label{wavefunction}

\subsection{Inclusive Deep Inelastic Scattering} \label{dis}

Our most precise knowledge of nucleon (and nuclear) structure is based on
deep inelastic lepton scattering (DIS), $\ell N \to \ell' X$, and related hard
inclusive reactions. As is well-known, DIS measures the product of a
parton-level subprocess cross-section $\hat \sigma$ and a target structure
function $F$. Thus, schematically and at lowest order in the strong coupling
$\as$,
\beq
\frac{d^2\sigma(eN\to eX)}{dQ^2dx} = \hat\sigma(eq\to eq)\, F_{q/N}(x,Q^2)\,
[1+\morder{\as}]  \label{epex}
\eeq
The structure functions $F_{q/N}$ have been measured over an impressive range
in $x$ and $Q^2$, covering $.0001 \lsim x \lsim 1$ and $1 \lsim Q^2 \lsim
10000\ \gev^2$. Their logarithmic $Q^2$-dependence (`scaling violations')
predicted by QCD has been tested, and their `universality' verified, \ie, the
same structure functions describe other hard inclusive reactions such as $pp
\to jet+X$, $\pi^-p \to \mu^+\mu^- + X$ and $pp \to \gamma + X$. The many
detailed measurements and successful cross-checks have together established
QCD as the correct theory of the strong interactions, and made us confident
that basic properties of hadron wave functions can be deduced from experimental
measurements using the methods of PQCD.

The success of DIS phenomenology should not make us forget that the
structure function $F_{q/N}(x,Q^2)$, no matter how completely known, still
only provides a very limited knowledge of the nucleon wave function. In
terms of a (light-cone) Fock state expansion of the proton wave function,
\beqa
\ket{p}&=& \int\prod_i\, dx_i\, d^2k_{\perp i} \left\{
\Psi_{uud}(x_i,k_{\perp i}) \ket{uud} \right. \nonumber \\
&+& \left. \Psi_{uudg}(\ldots) \ket{uudg}+ \ldots + \Psi_{\cdots}(\ldots)
\ket{uudq\bar q}+ \ldots \right\} \label{fock}
\eeqa
the structure function $F_{q/p}$ can be expressed as a sum over the absolute
squares of all Fock components $n$ that contain a parton $q$ with the measured
momentum fraction $x$,
\beq
F_{q/p}(x,Q^2)= \sum_n \int^{k_{\perp}^2<Q^2} \prod_i\, dx_i\, d^2k_{\perp i}
|\Psi_n(x_i,k_{\perp i})|^2 \delta(x-x_q)  \label{strfn}
\eeq
Due to the average over Fock states, the most probable states will typically
dominate in the structure function. Information about partons which do not
participate in the hard scattering is lost in the sum of \eq{strfn}. The
structure function is a single parton inclusive probability distribution that
does not teach us about parton correlations. However, at large values of
$x$ the structure function singles out unusual Fock states where one parton
carries nearly all momentum, and all other partons therefore must have low $x$.

\subsection{Hard Exclusive Scattering} \label{exclusive}

It is desirable to make measurements of hadron wave functions beyond the
structure function (\ref{strfn}). This requires special care, given
that we only master the perturbative region of QCD. We need to study a hard
scattering, where the subprocess can be identified and calculated, and where
the dependence on the soft wave function factorizes. The factorization
between hard and soft processes is a nontrivial feature in a theory like QCD
with massless (long-range) gluon exchange. Even in inclusive scattering
factorization has only been proved for a subset of the measureable hard
processes \cite{fact}.

\begin{figure}[htb]
\begin{center}
\leavevmode
{\epsfxsize=13.5truecm \epsfbox{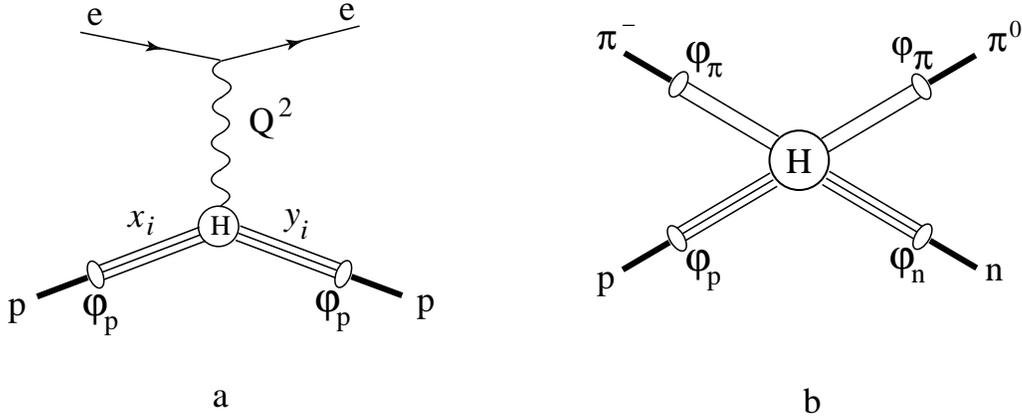}}
\end{center}
\caption[*]{a. Elastic $ep\to ep$ scattering at large $Q^2$ factorizes into a
product of proton distribution amplitudes $\varphi_p$ and a hard electron
scattering from the compact valence Fock state $\ket{uud}$. b. An analogous
factorization is illustrated for the large angle process $\pi^-p \to \pi^0n$.}
\label{eptoep}
\end{figure}

The hard subprocess can occur coherently off several partons if the distance
between them is commensurate with the momentum transfer $Q$. Such (`higher
twist') processes are more strongly damped in momentum transfer than DIS
(\cf\ Eqs. (\ref{siginc}) and (\ref{sigexc})), since the partons must be
increasingly close as $Q$ grows. This is the case in hard exclusive
processes, where factorization has also been shown to apply \cite{brle}.
As an example, consider elastic electron-proton scattering, $ep\to
ep$ at large momentum transfer $Q$ (Fig. \ref{eptoep}a). The amplitude for this
process factorizes into a product of a hard
scattering part $T_H$ and proton `distribution amplitudes' $\varphi_p$,
\beq
A(ep\to ep) = \int_0^1 \prod_{i=1}^3 dx_i dy_i \varphi_p(x_i,Q^2)\, T_H\,
\varphi_p(y_i,Q^2) \{1+\morder{1/Q^2}\}  \label{epep}
\eeq

The proton distribution amplitude is the valence part of the Fock expansion
(\ref{fock}), integrated over relative transverse momenta up to $Q$,
\beq
\varphi_p(x_i,Q^2) = \int^{k_{i\perp}^2 < Q^2} \prod_i d^2\vec k_{i\perp}
\Psi_{uud}(x_i,k_{i\perp})  \label{distamp}
\eeq
where the $x_i$ denote the longitudinal momentum fractions of the valence
$uud$ constituents. The integral over the relative transverse momenta
$k_{i\perp}$ implies that the transverse size of the valence state is
$r_\perp \simeq 1/Q$. The hard amplitude $T_H$ describes the subprocess
$e+(uud) \to e+(uud)$, which selects compact $\ket{uud}$ states.

The logarithmic $Q^2$ dependence of the proton distribution amplitude is given
by
\beq
\varphi_p(x_i,Q^2)=120 x_1 x_2 x_3 \delta(1-x_1-x_2-x_3) \sum_{n=0} \left[
\frac{\as(Q^2)}{\as(Q_0^2)} \right]^{\lambda_n} C_n P_n(x_i)
\label{qdep}
\eeq
The anomalous dimensions form an increasing series
\beq
\lambda_0= \frac{2}{27} < \lambda_1= \frac{20}{81} < \lambda_2= \frac{24}{81} <
 \ldots
\eeq
implying that each successive term in \eq{qdep} decreases faster with $Q^2$
than the previous one. The $P_n$ are Appell polynomials, $P_0=1$, 
$P_1=x_1-x_3$, $P_2= 1-3x_2,\ \ldots$ and the $C_n$ are constants which
characterize the proton wave function and have to be determined from
experiment. The $Q^2$ evolution of the pion distribution amplitude is given by
an expression similar to \eq{qdep}. The overall normalization of the
pion distribution amplitude is fixed by the decay constant $f_\pi$ measured in
$\pi \to \mu\nu$ decay.

Just as in the case of inclusive scattering, the relevance of
factorization for data on exclusive reactions must be demonstrated by showing
that the same (universal) distribution amplitudes $\varphi_h$ describe
several hard exclusive processes. For example, large angle $\pi^-p \to \pi^0 n$
scattering should be described by the diagram of Fig. \ref{eptoep}b, which
involves the pion and proton distribution amplitudes and the $(q\bar
q)+(qqq)$ elastic subprocess.  Heavy meson decays like $B
\to \pi\pi$ can also be analyzed in the same formalism (assuming that the
momentum transfers involved are large enough).

Tests of factorization in exclusive reactions are quite difficult in
practice. From a theoretical point of view, the calculation of
multi-parton scattering amplitudes like those in Fig. \ref{eptoep} are very
demanding even at the Born level, due to the large number of Feynman
diagrams. It is also difficult to estimate how high momentum
transfers are required in order to reach the scaling regime. Thus in Fig.
\ref{eptoep}a the momentum transfer $Q$ from the electron is effectively
split among the three quarks of the proton. The less momentum a quark
carries, the less transfer it needs to scatter to a large angle. There is
an especially dangerous region where some of the valence quarks carry a
very small fraction $x$ of the proton momentum, in which case they can
fit into the proton wave function both before and after the hard
scattering, without receiving any momentum transfer. There has been much
discussion as to the importance of this `Feynman mechanism' \cite{isgll}. The
consensus appears to be that it is suppressed asymptotically
\cite{brle,bost} due to the Sudakov effect \cite{suda}: The single quark
carrying all the momentum cannot be deflected to a large angle without gluon
emission. At finite (and relevant) energies, the importance of the Feynman
mechanism is still not settled -- and its significance may depend on the
reaction.

An immediate consequence of factorization for exclusive reactions is the
`counting' or 'dimensional scaling' rule \cite{dsr}, which gives the power of
the squared momentum transfer $t$ by which any $2\to 2$ fixed angle
differential cross section is suppressed (up to logarithms),
\beq
\frac{d\sigma}{dt}(2\to 2) \propto \frac{f(t/s)}{t^{n-2}}  \label{scarul}
\eeq
where $n$ is the total number of elementary fields (quarks, gluons, photons)
that are involved in the scattering. This rule follows from simple
geometrical considerations. Elastic scattering between two elementary
fields (\eg, $qq\to qq$) involves no dimensionful quantities except $s$ and
$t$ and thus obeys \eq{scarul} with $n=4$ at fixed $t/s$. Each additional field
that is involved in the scattering must be within a transverse distance of
order $r_\perp \lsim 1/Q$ (with $Q^2 \simeq -t$) to scatter coherently, and
the probability for that is of
\order{1/(Q^2 R^2)}, where $R \simeq 1$ fm is the average radius of the
hadron. This rule also explains why the dominant contribution to hard
scattering comes from the valence Fock states, which minimize the power $n$
in \eq{scarul}.

It is encouraging (although by no means conclusive) for factorization in
hard exlusive processes that the scaling rule (\ref{scarul}) is
approximately obeyed by the data for many reactions. Thus, $ep \to ep$
involves a minimum of $n=8$ fields, implying that the
proton form factor should scale as $F_p(Q^2) \propto 1/Q^4$, as assumed in
\eq{sigexc}. Data is available
\cite{pform} for $Q^2 \lsim 30\ \gev^2$ and is consistent with this behavior
for $Q^2 \gsim 5\ \gev^2$. At the higher values of $Q^2$ there are indications
of scaling violations that are consistent with the logarithmic evolution
predicted by \eq{epep}.

Tests of the dimensional scaling rules in exclusive reactions are
analogous to tests of Bjorken scaling in DIS, \ie, that the $Q^2$
dependence of the inclusive cross section is given by \eq{siginc}. In DIS,
the cross section as a function of $x$ then directly measures the structure
function $F(x)$. In exclusive reactions the situation is not as favorable.
The experimentally determined normalization of the proton form factor only
gives us one number, which is an average of the proton
distribution amplitude integrated over the momentum fractions $x_i$ carried
by the valence quarks. To make a quantitative prediction one must know both
the shape and the normalization of the (non-perturbative) distribution
amplitude. The good news is that the asymptotic form of the amplitude in the
$Q^2\to\infty$ limit is known, $\varphi_p^{AS} \propto x_1x_2x_3$ according to
\eq{qdep}. The non-asymptotic corrections are encoded in the moments $C_i$
which are measurable in principle. Considerable efforts have been made
to determine the pion and proton distribution amplitudes theoretically using
lattice calculations and QCD sum rules \cite{sumrules,rads}.

\begin{figure}[htb]
\begin{center}
\leavevmode
{\epsfxsize=13.5truecm \epsfbox{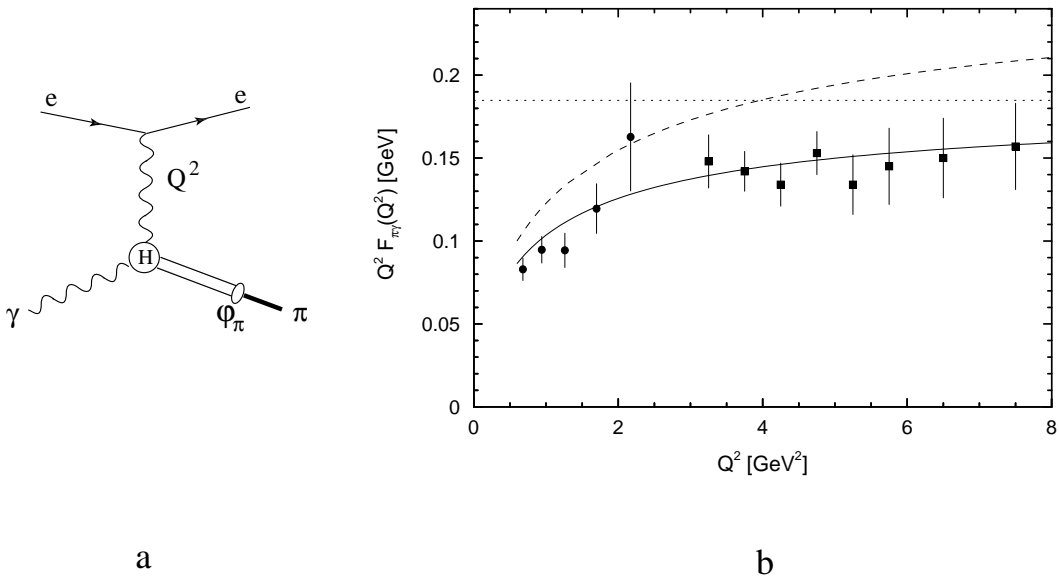}}
\end{center}
\caption[*]{a. The pion transition form factor $F_{\pi\gamma}$ is measured
by the process $e\gamma \to e\pi$, and factorizes at high $Q^2$ into a
product of the calculable hard subprocess $e\gamma \to e+(q\bar q)$ and the
pion distribution amplitude $\varphi_\pi$. b. Data \cite{pida}
compared with calculations based on a pion distribution amplitude
close to the asymptotic one (solid line) and one based on QCD sum rules
\cite{sumrules} (dashed line). The dotted line represents the asymptotic
result $\sqrt{2}f_\pi$. Figure from Kroll \etal\ in \cite{pitff}.}
\label{fpigam}
\end{figure}

One of the simplest hard exclusive processes is the pion transition form
factor $F_{\pi\gamma}(Q)$, measured by the process $e\gamma \to e\pi$ at large
momentum transfer $Q$, \cf\ Fig. \ref{fpigam}a. The existing data \cite{pida}
in the range
$1<Q^2<8\ \gev^2$ shown in Fig. \ref{fpigam}b is well fit using a pion
distribution amplitude close to the asymptotic form $\varphi_{\pi}^{AS}
\propto x_1x_2$ \cite{pitff}. Considering that the absolute normalization in
the large $Q^2$ limit is fixed by the pion decay constant,
$F_{\pi\gamma}^{AS}=\sqrt{2}f_\pi/Q^2$, the agreement is very encouraging and
indicates that the factorization formalism applies even at moderate
values of $Q^2$. There is evidence, on the other hand, that the asymptotic
regime may be more distant in the case of the pion form factor measured by
$e\pi \to e\pi$ large angle scattering \cite{piff}.

There are many other processes that can and need to be
analyzed experimentally and theoretically in order to achieve a
comprehensive understanding of the phenomenology of hard exclusive
scattering. A particularly important process is virtual compton scattering
$\gamma^*p\to \gamma p$, which involves no hadrons except the proton and
offers the possibility of varying independently both the virtuality of the
photon and the momentum transfer to the proton \cite{shupe,krni,rads,ji}. Many
exclusive processes involve resonance production and thus require the
measurement of multiparticle final states. It seems clear that the
phenomenology of rare exclusive processes requires the capability of an \elf\ 
type accelerator, which combines sufficient energy with high luminosity in a
continuous electron beam.

\subsection{Scattering from Compact Subsystems} \label{subcompact}

In inclusive DIS and in hard exclusive processes a photon (or gluon)
scatters from a parton system ($q, g, q\bar q$ or $qqq$) with a transverse size
of \order{1/Q}, compatible with the photon wavelength. Intuitively, this is
required for the physics of the hard perturbative scattering to factorize from
the non-perturbative wave function, which determines the probability for such
compact systems.

\begin{figure}[htb]
\begin{center}
\leavevmode
{\epsfxsize=13.5truecm \epsfbox{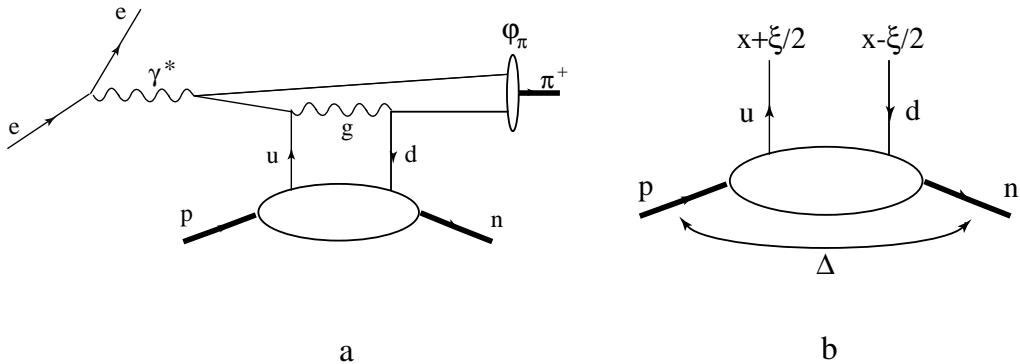}}
\end{center}
\caption[*]{a. Lowest order contribution to the process $ep \to e\pi^+n$, in
the limit of large $\nu$ and $Q^2$ but a fixed momentum transfer
between the nucleons. b. An off-forward parton distribution, with the
momentum fractions $x\pm\xi/2$ of the quarks and the momentum transfer
$\Delta$ between the nucleons indicated.} 
\label{gamptopin} 
\end{figure}

Fully inclusive scattering like DIS measures single parton distributions, with
no constraint on the size of the Fock state to which they belong. In exclusive
scattering the whole Fock state is required to be compact. There are also
hard processes where the scattering occurs off
{\em multiparton subsystems} of the hadron, such as
$qq,\ gg,\ etc.$ The theoretical framework for such processes is still
incomplete, but quite recently progress has been made for meson production
processes such as $\gamma^* p \to \pi^+n$ shown in Fig. \ref{gamptopin}a. In
the limit where the energy $\nu$ and virtuality $Q^2$ of the photon are large,
but the momentum transfer $\Delta$ between the nucleons remains fixed, the
amplitude for this process factorizes \cite{colfrastr} into a perturbatively
calculable hard subprocess (shown to lowest order in Fig. \ref{gamptopin}a),
the distribution amplitude $\varphi_\pi$ of the $\pi^+$ and an `Off-Forward
Parton Distribution' (OFPD) shown in Fig. \ref{gamptopin}b. As suggested by
the figure, the OFPD is a generalization of the usual quark structure
function measured in DIS to the non-forward $(\Delta,\ \xi \neq 0)$ and
inelastic $(p \neq n)$ case.

It has been shown \cite{rads,ji,dvcs} that the OFPD's interpolate between
structure functions and distribution amplitudes. In the kinematic region where
the light-cone momentum fractions $x\pm\xi/2$ of the quarks in Fig.
\ref{gamptopin}b are both positive (or negative) the OFPD is analogous to
a structure function. On the other hand, when for example $x-\xi/2<0$, this
vertex should be regarded as the distribution amplitude of a compact $u\bar
d$ pair in the nucleon, which carries momentum fraction $\xi$. In this case
the physics corresponds to the virtual photon scattering off the quark pair and
ejecting it from the nucleon, where it forms a $\pi^+$. The $\gamma^* p \to
\pi^+n$ process thus can give experimental information on the distribution of
compact quark pairs in the nucleon.

\begin{figure}[htb]
\begin{center}
\leavevmode
{\epsfxsize=13.5truecm \epsfbox{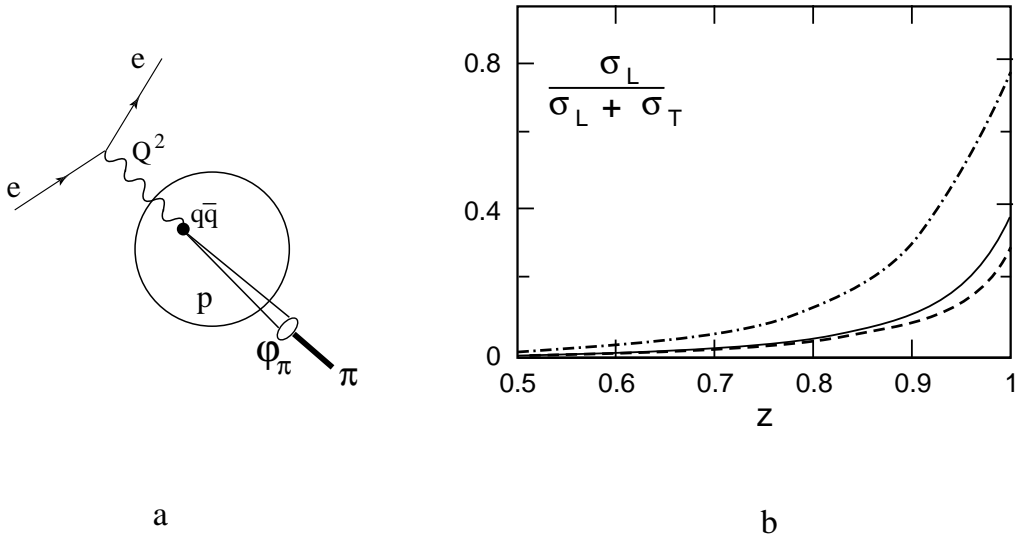}}
\end{center}
\caption[*]{a. Electron scattering off compact $\qpair$ pairs in the target
are selected by the semi-inclusive process $ep\to e\pi+X$ when the pion
carries a large fraction $z$ of the photon energy. b. Model
calculation \cite{bra} of the ratio $\sigma_L/(\sigma_L+\sigma_T)$, showing
how coherent scattering on $\qpair$ begins to dominate at large $z$. The
curves correspond to different choices of the pion distribution amplitude.}
\label{gamtopix}
\end{figure}

Consider now the semi-inclusive process $ep \to
e\pi + X$ sketched in Fig. \ref{gamtopix}a, where the pion takes a fraction
$z$ of the photon energy $\nu$. In the limit $z\to 1$ the photon transfers
nearly all its energy to the pion. This requirement selects compact $\qpair$
configurations \cite{bb,bra}. Alternatively (and in fact equivalently), the
struck quark needs to combine with a very soft antiquark to form the pion.
Such asymmetric configurations are short-lived and indistinguishable (by the
photon) from compact $q\bar q$ pairs in the limit of $z \to
1$ with fixed $(1-z)Q^2$ \cite{bhmt}. Although a formal proof is still
lacking, we may thus expect the cross section to factorize in this limit as
\beq
\sigma= F_{\qpair/p}(x)\,\hat\sigma(e+(\qpair) \to e+(\qpair))\,
|\varphi_\pi|^2 \label{piprod}
\eeq
where $F_{\qpair/p}(x)$ is the probability for finding the compact quark
pair in the target, and the pion distribution amplitude $\varphi_\pi$ is
the amplitude for the pair to transform into a physical pion.

Scattering off $q\bar q$ pairs (having integer spin) can be distinguished
from scattering off single (spin 1/2) quarks through the ratio
$\sigma_L/(\sigma_L+\sigma_T)$ of the longitudinally polarized to total
photon cross sections. As is well known, $\sigma_L=0$ (up to higher order QCD
corrections) for scattering from spin 1/2 quarks, whereas $\sigma_T=0$
for scattering on spin 0 diquarks. A calculation of the cross section ratio as
a function of $z$ based on the model orginally proposed in Ref. \cite{bb} is
shown in Fig. \ref{gamtopix}b. Experimental evidence for an analogous effect
has been seen in the reverse reaction $\pi N\to \mu^+\mu^- + X$, where the muon
pair takes a high fraction $x_F$ of the pion momentum \cite{e615}.

\begin{figure}[htb]
\begin{center}
\leavevmode
{\epsfxsize=13.5truecm \epsfbox{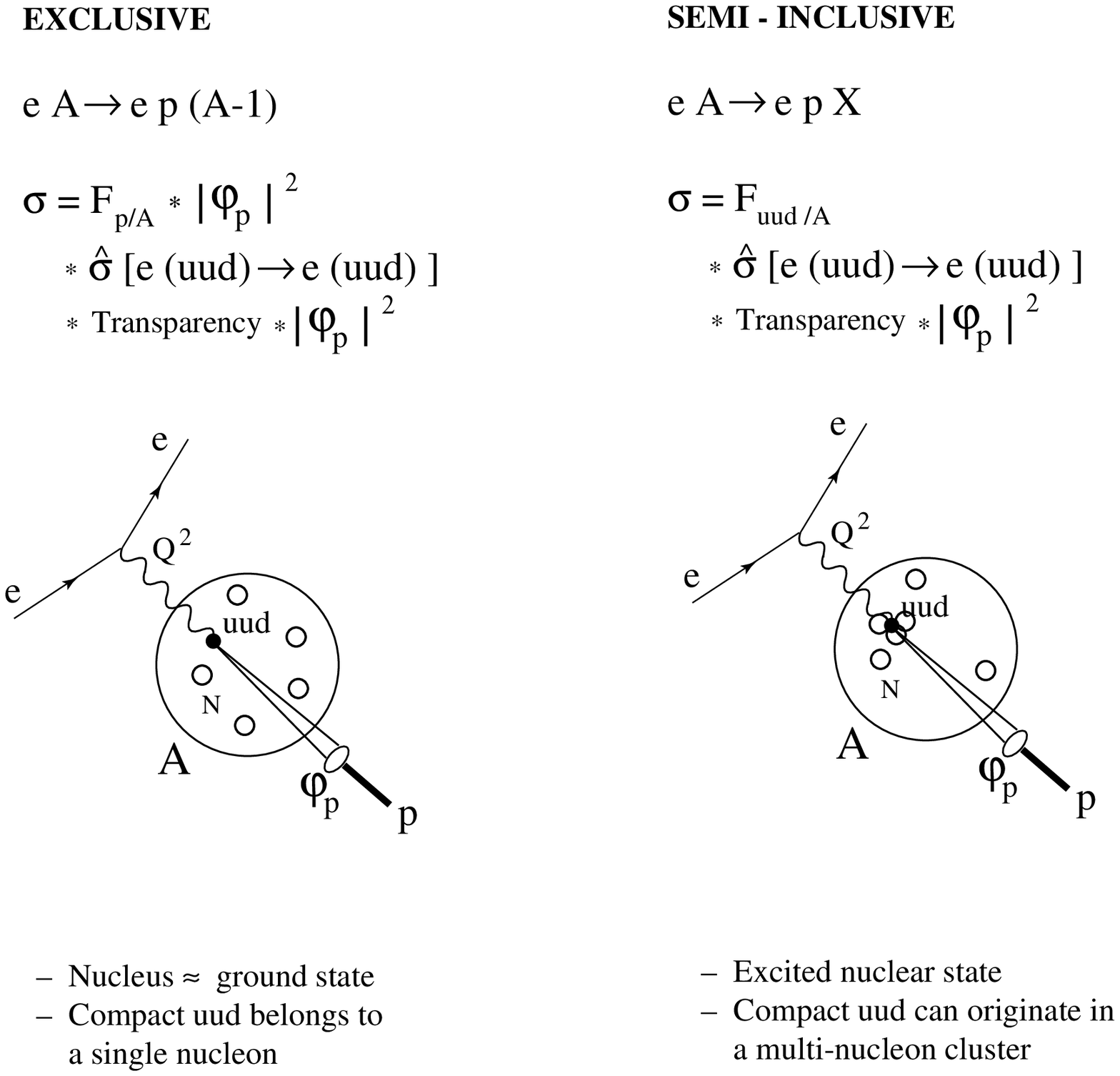}}
\end{center}
\caption[*]{Comparison of proton knock-out from nuclei in the exclusive
$eA \to ep(A-1)$ process and a semi-inclusive one, $eA \to epX$ with
$z=E_p/\nu \to 1$. The latter can have contributions from highly excited
nuclear states containing compact multinuclon clusters.}  
\label{gamtopx} 
\end{figure}

A factorization of the form (\ref{piprod}) for semi-inclusive processes
allows an interesting extension of the well-known color transparency
tests \cite{ct} using proton knock-out from nuclei. In the exclusive reaction
$eA \to ep(A-1)$ measured by NE-18 \cite{ne18} (and in $pA \to pp(A-1)$ as
measured at BNL \cite{bnlct}) the requirement that the the final state nucleus
is close to its ground state (no pion emission) selects `typical' nuclear
configurations. The probability to find a compact $uud$ state is then given by
the number $F_{p/A}=Z$ of protons in the nucleus multiplied by the
probability to find the $uud$ in a free proton (see Fig. \ref{gamtopx}). The
semi-inclusive reaction $eA \to epX$ with $1-z \propto 1/Q^2$, on the other
hand, can get contributions from highly excited nuclear configuration. The
quarks in the compact $uud$ state could, for example originate from separate
but overlapping nucleons. Comparing proton knock-out in exclusive and
semi-inclusive processes will thus tell us whether the latter contributions are
significant, 
\beq
F_{uud/A} {\buildrel ? \over =} Z\, |\varphi_p|^2  \label{uudtest}
\eeq
and hence will measure the distribution of compact multinucleon clusters in
nuclei.

\section{Short Range Correlations in Nuclei} \label{srcina}

The inclusive nuclear structure function is to a first approximation
given by the nucleon one, $F_{q/A}(x) \simeq A F_{q/N}(x)$ \cite{arn}.
Deviations of
\order{20 \ldots 30\ \%} are observed for small values of $x$ (`shadowing') and
for $x=0.5 \ldots 0.7$ (the `EMC effect'). When viewed in coordinate space, one
finds \cite{hova} that the quark `mobility distribution' is almost independent
(at the 2\% level) of $A$ up to light-cone distances (conjugate to $Q^2/2\nu$)
of order 2 fm, with shadowing setting in at larger distances. Since DIS is
dominated by the most common Fock states this indicates that typical
nucleon configurations are little affected by the nuclear environment. The
shadowing effect at large light-cone distances reflects coherent scattering
off several nucleons in the nucleus.

In contrast to inclusive scattering, hard semi-inclusive and exclusive
scattering select rare parton configurations, where some or all of the
partons in the Fock state are at short relative transverse distance. Since
such configurations do not contribute to DIS at moderate values of $x$ their
$A$-dependence is essentially unknown. Clusters that carry more momentum than
single nucleons in the nucleus are of special interest, since they select
nuclear configurations where several nucleons are at short relative distance.
In the parlance of nuclear physics, these represent highly excited states of
the nucleus (with excitation energies in the GeV region) about which we know
very little at present. An electron beam of high intensity and resolution is
essential for mapping out such dense clusters.

In DIS on nuclei, the fraction $x=Q^2/2m_p\nu$ of the target momentum carried
by the struck quark has the range $0 \leq x \leq A$. Data at $x \gsim
1$ exists and is difficult to explain by standard Fermi
motion \cite{bcdms,arrington,nutev}. Models based on short-range correlations
between nucleons \cite{frastri1} and on multi-quark effects \cite{bags} can fit
the data, but considerably more experimental and theoretical effort
will be needed to clarify the physics of this `cumulative' region of nuclei.

Novel cumulative effects are observed also in nuclear
fragmentation into hadrons \cite{frastri1,stav,geag}. The hadron ($p,\ \pi,\
K$) momentum distributions extend beyond $x_F=1$, \ie, their momentum must have
been transferred from several nucleons. The fragmentation is only weakly
dependent on the nature of the projectile or its energy, indicating that it
measures features intrinsic to the nuclear wave function. In these processes the
projectile scattering is soft, but there is evidence \cite{boya} that the
average transverse momentum of the produced hadrons increases with $x_F$,
reaching $\langle p_\perp^2 \rangle = 2\ \gev^2$ at $x_F=4$ for protons. The
cumulative momentum transfers thus appear to originate in a transversally
compact region of the nucleus.

Cumulative nuclear effects have furthermore been observed in subthreshold
production of antiprotons and kaons \cite{subth}. The minimal projectile
energy required for the process $pp\to \bar p+X$ on free protons
at rest is 6.6 GeV. The kinematic limit for $pA\to \bar p+X$ on a heavy
nucleus at rest is only $3m_N\simeq 2.8$ GeV. This reaction has been observed
for $A$ = $^{63}$Cu down to $E_{lab}\simeq3$ GeV, very close to kinematic
threshold. Scattering on a single nucleon in the nucleus would at this energy
require a Fermi momentum of \order{800} MeV. While the $pA$ data can be fit
assuming such high Fermi momenta, this assumption leads to an underestimate of
subthreshold production in $AA$ collisions by about three orders of magnitude
\cite{fermom}. 

It is possible that the subthreshold production of $K$ and $\bar p$ on nuclei
involves the same compact multiparton clusters that are responsible for
scattering with $x>1$ and $x_F>1$, although this is far from clear at present.
A study of subthreshold production using lepton beams could be quite
informative, since the locality of the reaction can be tuned through the
virtuality of the exchanged photon. A further possibility to pin
down the reaction mechanism is provided by subthreshold production of charm.

\section{Charm production at ELFE} \label{charmprod}

\subsection{General remarks on charm(onium) production} \label{charmintro}

\elf\ will
operate in the region of charm $(\cpair)$ threshold, which in the case of
real photons is at $E_{\gamma}^{th} \simeq 8\ldots 12$ GeV for $\jpsi,\ldots,
D\bar D$ production on free nucleons. Charmonium production has proved to
be a very sensitive measure of reaction mechanisms, as evidenced by
order-of-magnitude discrepancies found between QCD models and data
\cite{schulrev,sansoni,mangano}. Furthermore, the suppression of charmonium
production in heavy ion collisions is widely discussed as a potential signal
for the formation of a quark-gluon plasma \cite{satz}. I shall
discuss some of the puzzles of charmonium production and
how photo- and electroproduction close to threshold can give important new
clues to production mechanisms as well as to hadron and nuclear structure
\cite{frastri}.

\begin{figure}[htb]
\begin{center}
\leavevmode
{\epsfxsize=13.5truecm \epsfbox{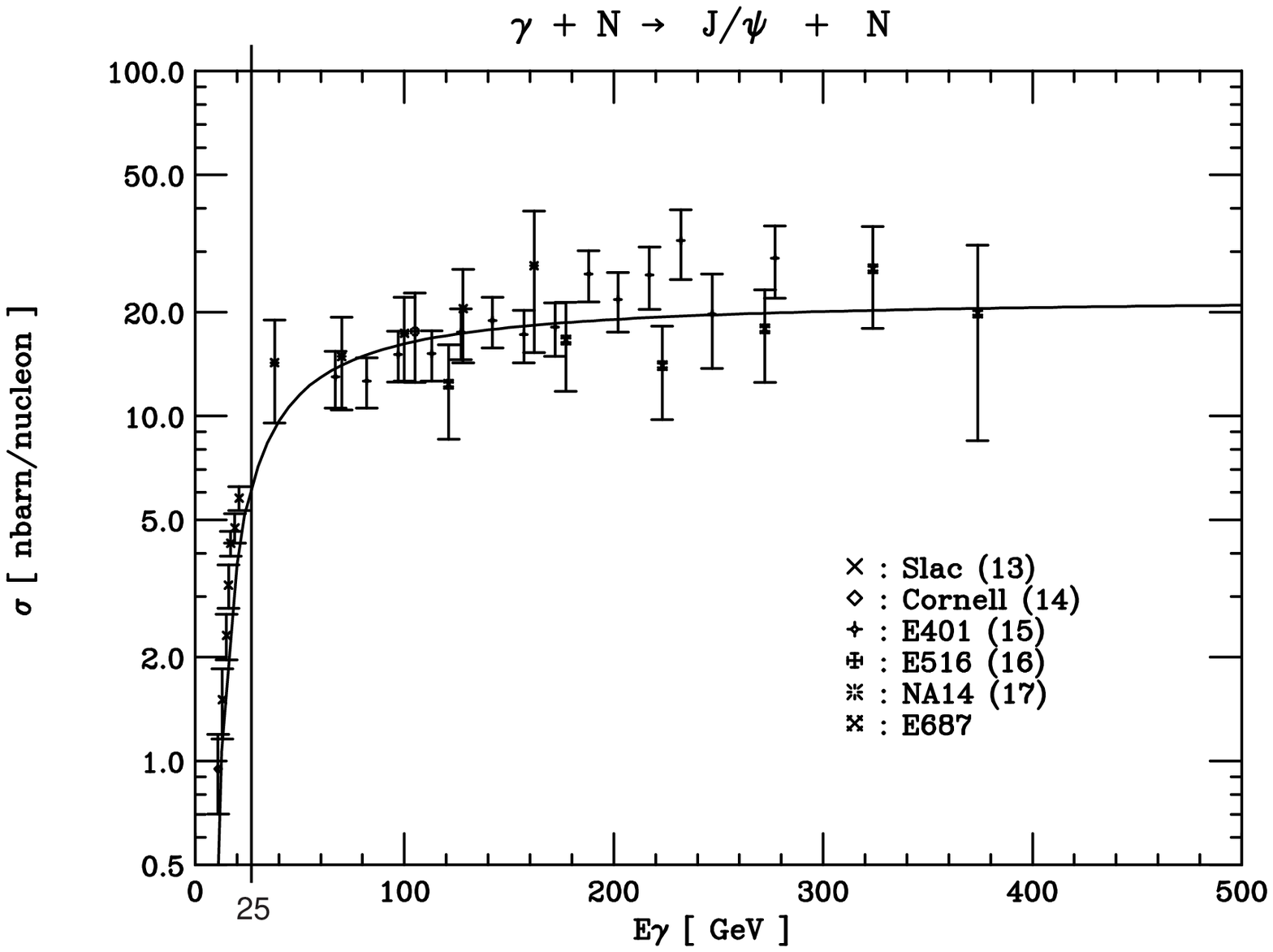}}
\end{center}
\caption[*]{Compilation of cross sections for the process $\gamma p\to \jpsi
p$ \cite{e687}. Experiments at ELFE will be in the range $E_\gamma\lsim 25$
GeV (vertical line). The curve shows the prediction of \eq{pgf} for a gluon
structure function $xG(x)= 3(1-x)^5$.}
\label{jpsilowel}
\end{figure}

Remarkably, the only charm photoproduction data that exists (Fig.
\ref{jpsilowel}) in the \elf\ energy range are the $\jpsi$ measurements of
SLAC \cite{slac} and Cornell \cite{cornell} from 1975, which predate the
discovery of open charm. These early measurements of the small near-threshold
cross section $\sigma(\gamma N \to \jpsi N) \simeq 1$ nb were made possible by
the experimental cleanliness of the $\jpsi \to \mu\mu$ signal. With an \elf\
luminosity ${\cal L} \sim 10^{35} {\rm cm^{-2}s^{-1}}$ one expects a rate of
about 5 $\jpsi$ dimuon decays per second, allowing detailed measurements of
threshold and subthreshold effects. 

It should also be kept in mind that owing to the
essentially non-relativistic nature of charmonium, each charm quark
carries close to one half of the $\jpsi$ momentum. Even their relative angular
momentum is determined through the quantum numbers of the charmonium state.
Charmonium is thus a very valuable complement to open charm
channels such as $D\bar D$, which furthermore are difficult to
measure.

Theoretically, charmonium offers very interesting challenges. Most reliable
QCD tests have so far concerned hard inclusive scattering, implying a sum
over a large number of open channels. The standard QCD
factorization theorem \cite{fact} does not apply when the final
state is restricted by requiring the charm quarks to bind as charmonium.
The application of QCD to charmonium production is thus partly an art, as
evidenced by lively discussions of different approaches. It seems likely
that charmonium production will teach us something qualitatively new about
QCD effects in hard scattering -- exactly what is not yet clear (but
hopefully will be so by the time \elf\ turns on!).

\subsection{$\jpsi$ production at high energies} \label{jpsihigh}

\subsubsection{Elastic $\jpsi$ Production} \label{eljpsi}

Early studies \cite{cem} assumed that the charmonium cross section is
proportional to the $\cpair$ one below open charm $(D\bar D)$ threshold, as
given by the inclusive photon-gluon fusion process $\gamma g \to \cpair$.
Thus
\beq
\sigma(\gamma N \to \jpsi+X) = f_{\jpsi} \int_{4m_c^2}^{4m_D^2} \frac{dM^2}{s}
G(M^2/s) \sigma_{\gamma g \to \cpair}\,(M^2)  \label{pgf}
\eeq
where $G(x)$ is the gluon structure function and the proportionality constant
$f_{\jpsi}$ is the fraction of the below-threshold $\cpair$ pairs that
form $\jpsi$'s. In this `Color Evaporation Model' (CEM) the
color exchanges which transform the color octet $\cpair$ pair into a
color singlet $\jpsi$ are assumed to occur over long time and distance
scales, and are described by the non-perturbative factor $f_{\jpsi}$ in
\eq{pgf}. For the model to have predictability it is important that this factor
be `universal', \ie, independent of the reaction kinematics (beam
energy and charmonium momentum), and hopefully also of the nature of the
projectile and target. It should be emphasized, however, that the universality
of $f_{\jpsi}$ is a hypothesis which has not been demonstrated in QCD.

Assuming a constant $f_{\jpsi}$ and a `standard' gluon structure function
$xG(x)=3(1-x)^5$, \eq{pgf} (with $X=N$) gives a good fit (solid line in Fig.
\ref{jpsilowel}) to $\jpsi$ elastic photoproduction from threshold to
$E_\gamma \lsim 300$ GeV \cite{e687}. It is not very clear what this means,
however. Close to threshold the single gluon exchange picture is expected to
break down (\cf\ section \ref{hte}). At high energy, color evaporation is
expected to apply to {\em inelastic} processes, since the neutralization of
color will lead to additional hadrons being produced.

\begin{figure}[htb]
\begin{center}
\leavevmode
{\epsfxsize=13.5truecm \epsfbox{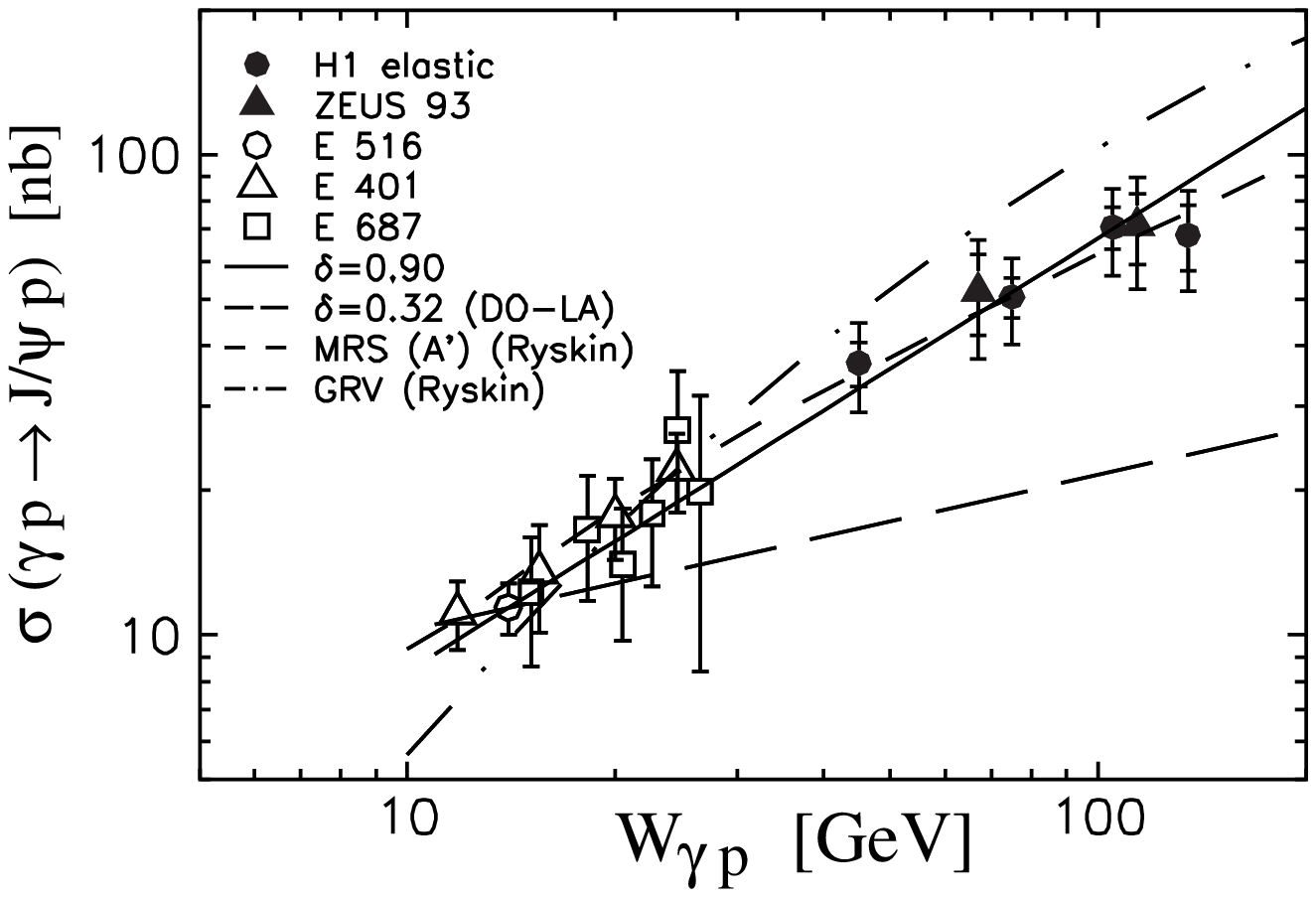}}
\end{center}
\caption[*]{Compilation of high energy data on the process $\gamma p\to \jpsi
p$ \cite{h1}, with curves of the form $W_{\gamma p}^\delta$ as indicated.
The curves marked `MRS' and `GRV' are the results of QCD calculations with
two-gluon exchange \cite{elqcd,rrml}, for different gluon structure functions.}
\label{jpsihighel}
\end{figure}

A consistent QCD description of high energy elastic $\jpsi$ photoproduction
involves two gluon (color singlet) exchange between the charm quark pair and
the target \cite{elqcd}. In this approach the color dynamics of the charm
quark pair is treated perturbatively, \ie, the quarks are created as a compact
color singlet state which couples directly to the $\jpsi$ through the wave
function at the origin. This is justified by the factorization between the hard
and soft physics in this process \cite{colfrastr}. As discussed in Sect.
\ref{subcompact} the two gluon coupling to the target
is an off-forward gluon distribution which near the forward direction may be
similar to the gluon structure function measured by deep inelastic
scattering \cite{dvcs}. The elastic $\jpsi$ cross section may thus be
approximately proportional to the {\em square} of the gluon structure
function. The high energy data (Fig. \ref{jpsihighel}) on $\gamma p \to \jpsi
p$ from HERA \cite{zeus,h1} in fact shows a considerable rise of the elastic
cross section with energy, which (within the considerable error bars) is
consistent with the increase of $xG(x)$ for $x \simeq 4m_c^2/s \to 0$
\cite{rrml}.

\subsubsection{Color Evaporation Approach to Inelastic $\jpsi$ Production}
\label{coleva}

The difficulties of perturbative QCD models in describing the data on
inelastic charmonium production (\cf\ sections 2.3 and 2.4 below) has
rekindled interest in the color evaporation model
\cite{gavai,schuler,amundson,schulvogt}. It has been shown that the dependence
of both charmonium and bottomonium production on the projectile energy and on
the energy fraction $x_F$ of the produced state are in good agreement with
that predicted through \eq{pgf} for heavy quarks below threshold. The fraction
of the below-threshold heavy quark cross section which ends up in quarkonium
depends on the QCD parametrization (quark mass, structure functions and
factorization scale) but seems to be quite small, typically (8 \ldots\ 10)\%
for charm, growing to (17 \ldots\ 32)\% for bottom \cite{schulvogt}. The
parameter $f_{\jpsi}$ of \eq{pgf} takes similar values in $pp$ and
$\pi p$ reactions (0.025 and 0.034, respectively \cite{gavai}). In
photoproduction the large diffractive (elastic) peak needs to be excluded, \eg,
by a cut on the $\jpsi$ momentum, after which values in the range $f_{\jpsi}=
0.005\ldots 0.025$ were found \cite{schulvogt}.

The generally good agreement of the color evaporation model with data is very
significant. It shows that the essential structure of the inclusive charmonium
cross section is given by that of heavy quark production at leading twist.
According to the spirit of color evaporation, the heavy quarks will after
their production undergo a long time-scale process of evolution to the
quarkonium bound state, during which the relative distance between the quarks
grows and non-perturbative gluons change the overall color of the
quark pair. The normalization of the production cross section, \ie, the
non-perturbative parameter $f_{\jpsi}$ in \eq{pgf}, is thus not necessarily
related to the wave function at the origin of the charmonium bound state.

\subsubsection{The Color Singlet Model (CSM)}  \label{colsing}

The `Color Singlet Model' (CSM) \cite{csm} describes charmonium
production fully in terms of PQCD. The $\cpair$ is created with proper quantum
numbers to have an overlap with the charmonium state, measured by the
non-relativistic wave function at the origin. In particular, the pair has to
be a singlet of color. For inelastic $\jpsi$ photoproduction the lowest order
subprocess is $\gamma g \to \cpair g$, where the final gluon radiation
ensures that the charmonium is produced with an energy fraction (in the
target rest frame) $z=E_{\jpsi}/E_\gamma<1$. For production at large $p_\perp
\gg m_c$ higher order `fragmentation diagrams' actually give the leading
contribution \cite{frag}.

\begin{figure}[htb]
\begin{center}
\leavevmode
{\epsfxsize=13.5truecm \epsfbox{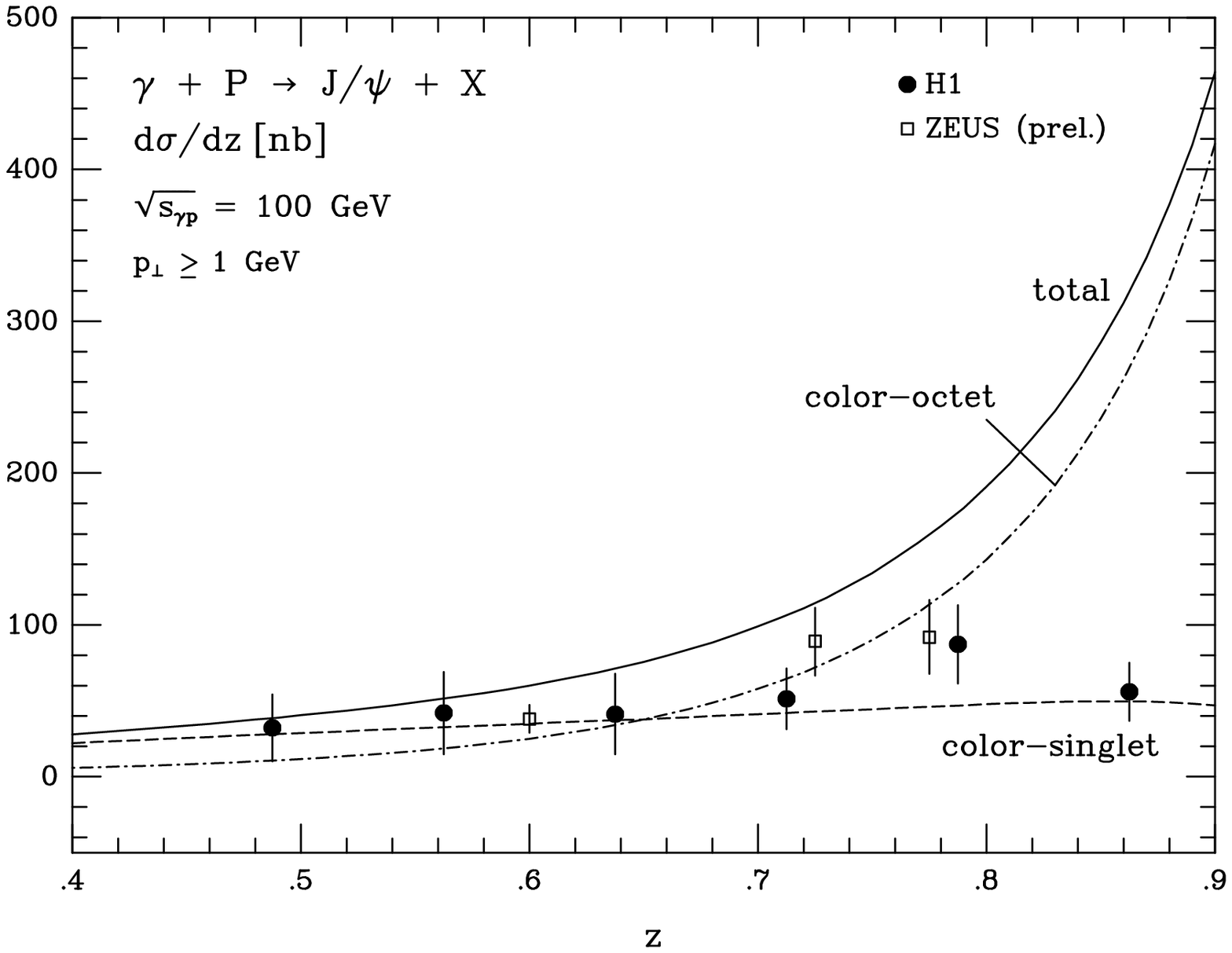}}
\end{center}
\caption[*]{The cross section for inelastic $\jpsi$ photoproduction $\gamma p
\to \jpsi+X$ for $p_\perp(\jpsi) \ge 1$ GeV as a function of the $\jpsi$ energy
fraction (in the proton rest frame) $z=E_{\jpsi}/E_\gamma$ \cite{cakr}.
Predictions based on the color singlet and octet mechanisms are compared to
data from HERA.}
\label{jpsiinel}
\end{figure}

The CSM contributions to $\jpsi$ photoproduction have been calculated to
next-to-leading order in QCD \cite{kramer}, with a result that is in good
agreement with the data (Fig. \ref{jpsiinel}). For $\psi'$ production the CSM
predicts that the $\psi'/\psi$ cross section ratio should be proportional to
the square of the wave function at the origin,
\beq
\frac{\sigma(\psi')}{\sigma_{dir}(\jpsi)}=\frac{\Gamma(\psi'\to
\mu\mu)}{M_{\psi'}^3}\frac{M_{\jpsi}^3}{\Gamma(\jpsi\to \mu\mu)}
 \simeq .24\pm .03  \label{dirratio}
\eeq
Here $\sigma_{dir}(\jpsi)$ excludes contributions to the $\jpsi$ from
`indirect' channels such as $B$, $\chi_c$ and $\psi'$ decays, and the power of
mass is motivated by dimensional arguments. The photoproduction data
\cite{na14,nmc} gives for the ratio that is uncorrected for radiative decays,
\beq
\frac{\sigma(\gamma N \to \psi'+X)}{\sigma(\gamma N \to \jpsi+X)}=0.20\pm 0.05
\pm 0.07  \label{fullratio}
\eeq
The upper limit on the $\chi_{c1}+\chi_{c2}$ photoproduction cross section is
about 40\% of the $\jpsi$ cross section \cite{na14}. Taking into account the
\order{20\%} branching ratio for their radiative decays into $\jpsi$ only a
small fraction of the photoproduced $\jpsi$'s are due to the indirect channels,
and the ratios of Eqs. (\ref{dirratio}) and (\ref{fullratio}) should be
compatible, as indeed they are. It should be noted, however, that the
experimental ratio (\ref{fullratio}) primarily reflects diffractive (elastic)
$\jpsi$ and $\psi'$ production, which dominates in photoproduction.

In hadroproduction, where inelastic channels dominate the cross section,
data on the ratio (\ref{dirratio}) is also in good agreement with the
color singlet model \cite{vhbt}. This is
true as well for the Tevatron data on charmonium production at large $p_\perp$
\cite{sansoni}. Even bottomonium production is quite consistent with the
analog of
\eq{dirratio}, within factor of two uncertainties due to the so far unmeasured
contributions from radiative decays of the P states \cite{gavai}.

The above comparisons suggest that the `nonperturbative' proportionality
factors $f$ in the color evaporation model (\cf\ \eq{pgf}) actually reflect
perturbative physics, \ie, the wave function at the origin as assumed in
the color singlet model.

In spite of its successful predictions in photoproduction and of the {\em
ratio} of $\psi'$ to $\jpsi$ hadroproduction, the CSM nevertheless
fails badly, by factors up to 30 \ldots\ 50, for the {\em
absolute hadroproduction} cross sections of the $\jpsi$, the $\psi'$ and
the $\chi_{c1}$ states \cite{schulrev,sansoni,mangano,vhbt}. The discrepancies
are large both in fixed target total cross section data and in
large $p_\perp$ production at the Tevatron. The fixed target data moreover
shows that the $\jpsi$ and $\psi'$ are produced nearly unpolarized
\cite{fixpol}, contrary to the CSM which predicts a fairly large transverse
polarization \cite{vhbt}.

The fact that the CSM {\em underestimates} charmonium hadroproduction (and
predicts the polarization incorrectly), suggests that there are other
important production mechanisms, beyond the CSM. The nature of those
processes is not yet established. A simple mnemonic, which appears to be
consistent with the observed systematics, is that the CSM works whenever no
extra gluon emission is required only to satisfy the quantum number
constraints. Thus, for inelastic $\jpsi$ photoproduction the lowest order
process $\gamma g \to \cpair g$ of the CSM has only the number of gluons which
is required by momentum transfer (and the prediction is successful). In the
(incorrect) CSM prediction for hadroproduction  gluon
emission in the subprocess $gg\to \cpair g$ is needed only due to the negative
charge conjugation of the $\jpsi$ (or due to Yang's theorem in the case of
$\chi_{c1}$ production). Again, for $\chi_{c2}$ the lowest order process $gg
\to \chi_{c2}$ is allowed in the CSM, and the prediction is compatible with
the data (within the considerable PQCD uncertainties)
\cite{mangano,vhbt,cgmp}. A polarization measurement of hadroproduced
$\chi_{c2}$'s would be a valuable check of the CSM \cite{vhbt,chipol}.

Photoproduction of $\chi_c$ is an interesting test case
\cite{schulvogt,cakr}. The available data \cite{na14} suggests that the
$\chi_{c2}/\jpsi$ ratio is lower in inelastic photoproduction than in
hadroproduction. This qualitatively agrees with the CSM, in which $P$-wave
photoproduction $(\gamma g\to \chi_{c2} gg)$ is of higher order than $S$-wave
production $(\gamma g\to \jpsi g)$, while the reverse is true for
hadroproduction. With no regard to quantum numbers (as in the color
evaporation model) the basic subprocess would be the same ($\gamma g
\to \cpair$) and the $\chi_{c}/\jpsi$ ratio would be expected to be similar
in photo- and hadroproduction.

It has been suggested that the gluons required to satisfy quantum number
constraints of the $\cpair$ pair in the CSM could come from additional (higher
twist) exchanges with the projectile or target \cite{vhbt,htwist}. Although
normally suppressed, these contributions might be important since they do not
involve energy loss through gluon emission.

\subsubsection{The Color Octet Model (COM)}  \label{coloct}

A possible solution to some of the above puzzles has been
suggested based on an analysis of nonrelativistic QCD (NRQCD) \cite{nrqcd},
and commonly referred to as the `Color Octet Model' (COM) \cite{cgmp,com}. In
cases where, due to quantum number constraints, extra gluon emission is
required in the CSM the production may be dominated instead by higher order
terms in the relativistic ($v/c$) expansion of the quarkonium bound state.
For $P$-wave states the inclusion of relativistic corrections is in fact
necessary to cancel infrared divergencies of the perturbative expansion even
at lowest order.

The $\cpair$ can then be produced in a color octet state, which has
an overlap with a higher $\ket{\cpair g}$ Fock state of charmonium, with the
emission of a soft gluon. Such contributions appear in a systematic NRQCD
expansion and thus must exist. Whether they are big enough to account for the
large discrepancies of the CSM in charmonium production depends on the
magnitude of certain non-perturbative matrix elements of NRQCD. I refer to
recent reviews \cite{mangano,annrev} and references therein to the
extensive literature on this subject.

A number of discrepancies between the color octet model and observations
suggest that it will at best provide only a partial explanation of quarkonium
production.
\begin{itemize}

\item[{\em (i)}] Inelastic photoproduction of $\jpsi$ is overestimated by the
COM \cite{cakr,photoprod} as seen in Fig. \ref{jpsiinel}. A best estimate of
the discrepancy is actually even larger than shown, since the effects of soft
gluon radiation were neglected in fitting the octet matrix elements from the
Tevatron data \cite{softg}. The photoproduction cross section can be {\em
decreased} in the COM only by adding a contribution which is coherent with
the production amplitude.

\item[{\em (ii)}] The COM does not explain the $p_\perp$-integrated (fixed
target) charmonium hadroproduction data, in particular not the polarization of
the $\jpsi$ and $\psi'$ and the $\chi_{c1}/\chi_{c2}$ ratio \cite{tv,bz,gs}.
It has been claimed, but is by no means obvious, that the (higher twist?)
corrections are bigger in the fixed target data than in the large $p_\perp$
cross section measured at the Tevatron, which is often taken as a benchmark
for COM fits. The systematics of the anomalies is actually very similar in
the two processes. The color singlet model fails by a comparable
factor in both cases \cite{sansoni}, while the (leading twist) color
evaporation model successfully explains the relative production rates measured
in the fixed target and Tevatron experiments \cite{amundson,schulvogt}.

\item[{\em (iii)}] The $\Upsilon(3S)$ cross section exceeds the CSM
predictions by an order of magnitude \cite{sansoni,cdfups}. Since the
relativistic corrections are much smaller for bottomonium than for charmonium,
this is hard to accomodate in the COM \cite{bz}. It has been suggested that
the excess could be due to radiative decay from a hitherto undiscovered $3P$
state. As in the case of the `$\psi'$ anomaly', an experimental measurement
of direct $\Upsilon$ production should settle this question. The ratios of
$\Upsilon(nS)$ cross sections are quite compatible with expectations based on
the wave function at the origin (\cf\ \eq{dirratio}), with only moderate
contributions from radiative decays of $P$ states \cite{gavai}. 
\end{itemize}

\subsubsection{Nuclear Target $A$-Dependence}  \label{nucladep}

Additional clues to quarkonium production dynamics is offered by data
on the nuclear target $A$-dependence \cite{hoyrev}. In the standard
parametrization
\beq
\sigma(A) \propto A^{\alpha}  \label{adep}
\eeq
one expects $\alpha \simeq 1$ for hard incoherent scattering, which is
additive on all nucleons in the target nucleus. This behavior is verified
with good precision for the Drell-Yan process of large-mass lepton pair
production \cite{dy} as well as for open charm ($D$ meson) production at low
$x_F$ \cite{dprod}. However, for $\jpsi$ and $\psi'$ hadroproduction $\alpha
\simeq 0.92 \pm .01$ for $.1 \lsim x_F \lsim .3$ \cite{cha}. This suppression may be
interpreted as a rescattering of the charm quark pair in the nucleus, with an
effective cross section of 7 mb \cite{satz} for conversion to open charm
production. Such rescattering will affect the quantum numbers of the
$\cpair$ pair, and should thus be considered in color singlet and octet
approaches. For the color evaporation model the target dependence shows that
the proportionality factor $f_{\jpsi}$ in \eq{pgf} is not universal for all
processes.

The nuclear suppression of $\jpsi$ and $\psi'$ production increases with
$x_F$, with $\alpha(x_F=.6) \simeq .8$ \cite{cha}. This effect, which breaks
leading twist factorization \cite{hvs}, may be due to intrinsic charm
\cite{ic,icpsi} and involve the scattering of low momentum valence quarks
\cite{bhmt}. Ascribing the effect to parton energy loss in the nucleus
requires the $\langle p_\perp \rangle$ in the rescattering to be unexpectedly
large \cite{bh,jr}. The dynamics of charmonium production at large $x_F$ is
analogous to the large $p_\perp$ Tevatron data due to the `trigger bias'
effect: In both cases the charmonium carries a large fraction of the momentum
of the fragmenting particle.

In inelastic (virtual) photoproduction a nuclear {\em enhancement} of $\jpsi$
production is observed,  $\alpha =1.05\pm 0.03$ for $x_F < 0.85$ and
$p_\perp^2 > 0.4\ \gev^2$ \cite{nmc,e691}. Contrary to hadroproduction, the
momentum distribution of $\jpsi$ photoproduction peaks at large $x_F$. Hence
an explanation in terms of energy loss is conceivable \cite{hkz}.

In the region of the coherent peak for $\jpsi$ photoproduction on nuclei at
very low $p_\perp$ there is an even stronger nuclear enhancement. E691
\cite{e691} finds $\alpha_{coh} = 1.40 \pm 0.06 \pm 0.04$, while NMC
\cite{nmc} gives $\alpha_{coh}=1.19\pm 0.02$. If the $\cpair$ pairs are
 compact enough not to suffer secondary interactions in the nucleus, one
expects $d\sigma_{coh}/dp_\perp^2) \propto A^2 \exp(-cA^{2/3}p_\perp^2)$ ($c$
being a constant). Hence $\alpha_{coh}=4/3$ for the $p_\perp$-integrated cross
section, in rough agreement with the data.

\medskip

As should be clear from the above, quarkonium production offers
interesting challenges, which are not fully met by any one proposed mechanism.
It seems likely that we are learning something about color dynamics that
cannot be accessed within the standard, fully inclusive formalism of PQCD.
Color exchanges to the $\cpair$ evidently take place in ways not adequately
described by the CSM. The importance of the NRQCD contributions (which surely
are present at some level) remains to be clarified, as does the assumption by
the color evaporation approach that charmonium production constitutes
a universal fraction of the $\cpair$ cross section below open charm threshold.

\subsection{Production Near Kinematic Threshold} \label{kinthr}

As noted in Sect. \ref{charmintro}, almost all experimental information on
charmonium production is at relatively high energy. While we may hope that
at least some of the puzzles discussed in the preceding section will be solved
in the near future, an understanding of production near threshold will
have to wait for a dedicated machine like ELFE. In the following I shall
discuss some generic features of (sub-)threshold charmonium production
related to the composite nature of the beam and/or target\footnote{
Calculations of higher order perturbative, leading twist effects in heavy quark
production near threshold may be found in Ref. \cite{thr}.} \cite{frastri}. It
is likely that charm production close to threshold will teach us new
physics, over and beyond what is now being learnt at higher energies.

\subsubsection{Higher Twist Effects} \label{hte}

At high energy the dominant contribution to hard processes comes from
`leading twist' diagrams, characterized by only one parton from each
colliding particle participating in the large momentum transfer $(Q)$
subprocess. Since the time scale of the hard collision is $1/Q$, only
partons within this transverse distance can affect the process. The
likelihood that two partons are found so close to each other is typically
proportional to the transvers area $1/Q^2$, which thus gives the suppression
of higher twist, multiparton contributions.

Close to the kinematic boundary the higher twist effects are enhanced,
however. Thus for $\gamma p \to \cpair p$ very near threshold, all the
partons of the proton have to transfer their energy to the charm quarks
within their creation time $1/m_c$, and must thus be within this transverse
distance from the $\cpair$ and from each other. The longitudinal momentum
transfer at threshold (in the proton rest frame) is $\simeq m_c$. Hence only
compact proton Fock states, with a radius equal to the compton wavelength of
the heavy quark, can contribute to charm production at threshold.

\begin{figure}[htb]
\begin{center}
\leavevmode
{\epsfxsize=13.5truecm \epsfbox{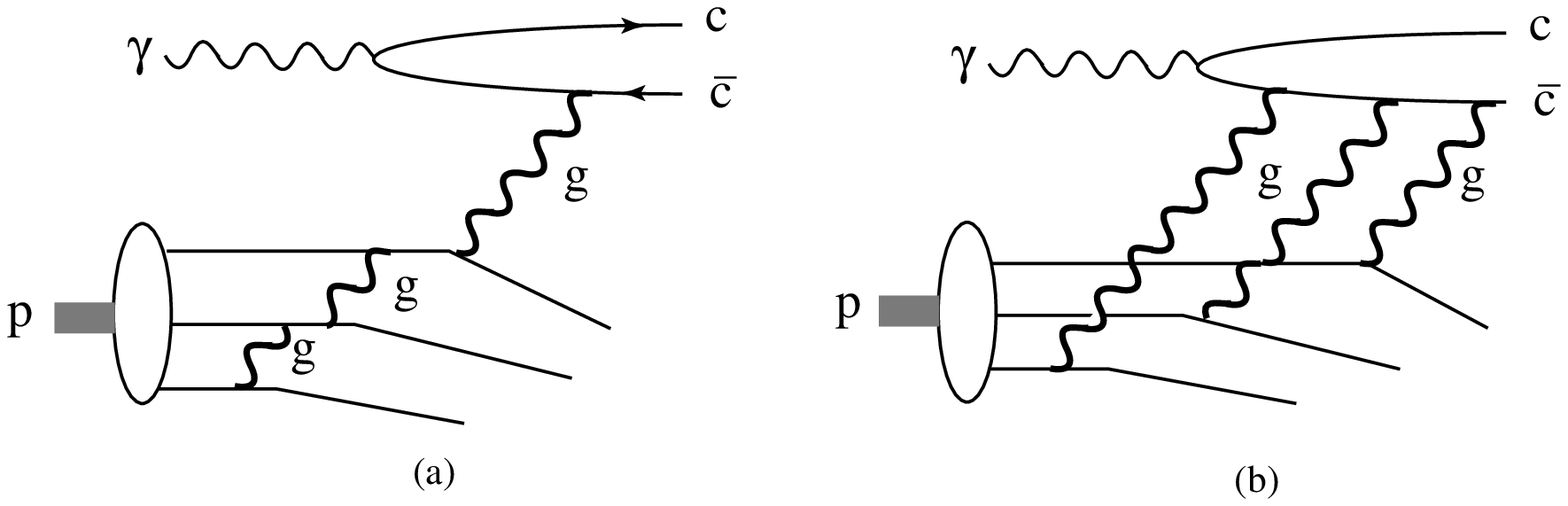}}
\end{center}
\caption[*]{Two mechanisms for transferring most of the proton momentum to the
charm quark pair in $\gamma p \to \cpair + X$ near kinematic threshold. The
leading twist contribution (a) dominates at high energies, but becomes
comparable to the higher twist contribution  (b) close to threshold.}
\label{cmech}
\end{figure}

The behavior of the effective proton radius in charm photoproduction near
threshold can be surmised from the following argument. As indicated in Fig.
\ref{cmech}a, one mechanism for charm production is that most of the proton
momentum is first transferred to one (valence) quark, followed by a hard
subprocess
$\gamma q \to \cpair q$. If the photon energy is $E_\gamma = \zeta
E_\gamma^{th}$, where $E_\gamma^{th}$ is the energy at kinematic threshold
($\zeta \gsim 1$), the valence quark must carry a fraction $x=1/\zeta$ of
the proton's (light-cone) momentum. The lifetime of such a Fock state
(in a light-cone or infinite momentum frame) is $\tau \simeq 1/\Delta E$,
where 
\beq
\Delta E= \frac{1}{2p}\left[m_p^2-\sum_i \frac{p_{i\perp}^2+m_i^2}{x_i}
\right] \simeq -\frac{\Lambda_{QCD}^2}{2p(1-x)}  \label{dele}
\eeq
For $x=1/\zeta$ close to unity such a short-lived fluctuation can be created
(as indicated in Fig. \ref{cmech}a) through momentum transfers from valence
proton states (where the momentum is divided evenly) having commensurate
lifetimes $\tau$, \ie, with
\beq
r_\perp^2 \simeq \frac{1}{p_\perp^2} \simeq
\frac{\zeta-1}{\Lambda_{QCD}^2} \label{psize}
\eeq
This effective proton size thus decreases towards threshold $(\zeta \to 1)$,
reaching $r_\perp \simeq 1/m_c$ at threshold, $\zeta-1 \simeq
\Lambda_{QCD}^2/m_c^2$.

As the lifetimes of the contributing proton Fock states
approach the time scale of the $\cpair$ creation process, the time ordering of
the gluon exchanges implied by Fig. \ref{cmech}a ceases to dominate
higher twist contributions such as that of Fig. \ref{cmech}b \cite{bhmt},
which are related to intrinsic charm \cite{ic}. There are in fact reasons to
expect that the latter diagrams give a dominant contribution to charmonium
production near threshold. First, there are many more such diagrams. Second,
they allow the final state proton to have a small transverse momentum (the
gluons need
$p_\perp \simeq m_c$ to couple effectively to the $\cpair$ pair, yet the
overall transfer can still be small in Fig. \ref{cmech}b). Third, with several
gluons coupling to the charm quark pair its quantum numbers can match those of
a given charmonium state without extra gluon emission.

The above discussion is generic, and does not indicate how close to threshold
the new effects actually manifest themselves. While more quantitative model
calculations certainly are called for, this question can only be
settled by experiment. It will be desirable to measure both the cross section
and polarization for several charmonium states, as well as for
open charm. At present, there are only tantalizing indications for novel
phenomena at charm threshold, namely:
\begin{itemize}
\item {\em Fast $\cpair$ pairs in the nucleon.}
The distribution of charm quarks in the nucleon, as measured by deep inelastic
lepton scattering, appears \cite{emc} to be anomalously large at high $x$,
indicating a higher twist intrinsic charm component \cite{ic}. An analogous
effect is suggested by the high $x_F$ values observed in $\pi N\to \jpsi+
\jpsi +X$ \cite{twopsi}. A proton Fock state containing charm quarks with a
large fraction of the momentum will enhance charm production close to
threshold. 

\item {\em $\jpsi$ polarization in $\pi^- p\to \jpsi+X$ for $x_F\to 1$.}
Only compact projectile $(\pi)$ Fock states contribute in the limit
where the $\jpsi$ carries almost all of the projectile momentum.
It may then be expected that the helicity of the $\jpsi$ equals the helicity of
the projectile, \ie, the $\jpsi$ should be longitudinally polarized. This
effect is observed both in the above reaction \cite{fixpol} and (as already
discussed in Sect. \ref{subcompact}) in $\pi N \to\mu^+\mu^- + X$ \cite{e615}.

\item {\em Polarization in $pp \to pp$ large angle scattering.} There is a
sudden change in the $A_{NN}$ polarization parameter close to charm threshold
for 90$^\circ$ scattering \cite{ann}. It has been suggested that this is due to
an intermediate state containing a $\cpair$ pair, which has low angular
momentum due to the small relative momenta of its constituents
\cite{broter}. This idea could be tested at ELFE by investigating correlations
between polarization effects in large angle compton scattering, $\gamma p \to
\gamma p$, and charm production ($\gamma p \to \cpair p$) near
threshold.

\item {\em Change in color transparency at charm threshold.} Intermediate
states with a charm quark pair could also give rise to the sudden decrease
in color transparency observed in $pA \to pp(A-1)$ close to charm threshold
\cite{bnlct}. Due to the low momentum of the constituents they expand to a
large transverse size within the nucleus, thus destroying transparency
\cite{broter}. Again, $e A$ reactions could provide important tests at
ELFE.
\end{itemize}

\subsubsection{Subthreshold production}  \label{subthres}

The high luminosities at ELFE will allow detailed studies of subthreshold
production of charm(onium). It is well established that antiprotons and kaons
are produced on nuclear targets at substantially lower energies than is
kinematically possible on free nucleons \cite{subth}. Thus the minimal
projectile energy required for the process $pp\to \bar p+X$ on free protons
at rest is 6.6 GeV, while the kinematic limit for $pA\to \bar p+X$ on a heavy
nucleus at rest is only $3m_N\simeq 2.8$ GeV. Antiproton production has been
observed in $p+\, ^{63}$Cu collisions down to $E_{lab}^p\simeq 3$ GeV, very
close to kinematic threshold. Scattering on a single nucleon in the nucleus
would at this energy require a fermi momentum of \order{800} MeV. While the
$pA$ data can be fit assuming such high Fermi momenta, this assumption leads
to an underestimate of subthreshold production in $AA$ collisions by about
three orders of magnitude \cite{fermom}.

\begin{figure}[htb]
\begin{center}
\leavevmode
{\epsfxsize=13.5truecm \epsfbox{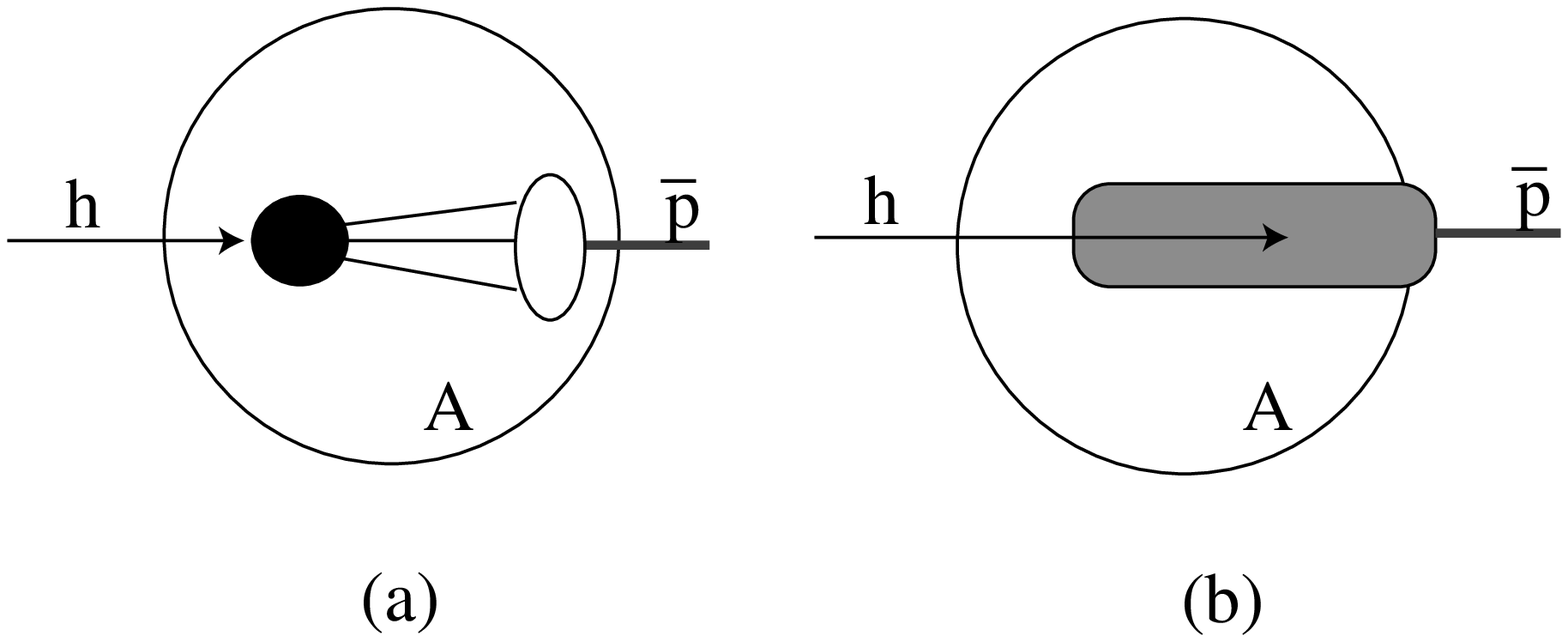}}
\end{center}
\caption[*]{Two conceptual mechanisms for subthreshold $\bar p$ production in
$hA$ collisions. In (a) the production occurs locally off a hot spot (black
circle) of high energy density in the nucleus. In (b) the light quarks gain
momentum over an extended nuclear region (grey).}
\label{submech}
\end{figure}

There are at least two qualitatively different scenarios for the observed
subthreshold production of antiprotons. Either (Fig. \ref{submech}a) the
projectile strikes a local `hot spot' with a high energy density in the
nucleus. The effective mass of the scatterer is high, lowering the
kinematic threshold. Alternatively (Fig. \ref{submech}b) the momentum required
to create the antiproton is not transferred locally, but picked up in an
extended longitudinal region: the nucleus forms a `femtoaccelerator'.
Establishing either scenario would teach us something qualitatively new about
rare, highly excited modes of the nucleus.

Real and virtual photoproduction of charm below threshold would be of crucial
help in distinguishing the correct reaction mechanism, for several reasons. 

\begin{itemize}
\item The photon is pointlike, and is thus a clean probe
of target substructure. In particular, effects due to the shrinking effective
size of a hadron probe near threshold (\cf\ discussion above) are eliminated.

\item The $\cpair$ pair is created locally, within a proper time $\tau \simeq
1/m_c$. The extended acceleration scenario of Fig. \ref{submech}b is thus not
effective for charm production. If significant subthreshold charm production
occurs (beyond what can be ascribed to standard fermi motion) this selects the
hot spot scenario of Fig. \ref{submech}a.

\item Subthreshold production can be studied as a function of the virtuality
$Q^2$ of the photon. Little $Q^2$ dependence is expected for $Q^2 \lsim
m_c^2$, due to the local nature of charm production. Nuclear hot spots
smaller than $1/m_c$ would be selected at higher values of $Q^2$.
\end{itemize}

\subsubsection{Interactions of $\cpair$ Pairs in Nuclei}  \label{ccina}

Close to threshold for the process $\gamma p \to \jpsi p$ on stationary
protons the energy of the $\jpsi$ is $E_{\jpsi}^{lab} \simeq 7$ GeV. This
corresponds to a moderate lorentz $\gamma$-factor $E_{\jpsi}/M_{\jpsi}
\simeq 2.3$. Hence a significant expansion of the $\cpair$ pair occurs
inside large nuclei, and effects of charmonium bound states in nuclei may be
explored.

Compared to the propagation of light quarks in nuclei, charm has the
advantage that one can readily distinguish hidden (charmonium) from open
$(D\bar D)$ charm production. Thus the dependence of the
$\sigma(\jpsi)/\sigma(D)$ ratio on the target size $A$ and on projectile
energy indicates the amount of rescattering in the nucleus. The
presently available data on the $A$-dependence of charmonium production is
at much higher energies (\cf\ Sect. \ref{nucladep}), and thus measures the
nuclear interactions of a compact $\cpair$ pair rather than of full-sized
charmonium. Further information about the significance of the radius of the
charmonium state can be obtained by comparing $\psi'$ to $\jpsi$ production on
various nuclei. In high energy $hA$ and $\gamma A$ scattering both states have
very similar $A$-dependence \cite{cha}.

Information about the propagation of charmonium in nuclei is very important
also for relativistic heavy ion collisions, where charmonium production may
be a signal for quark-gluon plasma formation \cite{satz}. Precise information
from ELFE would allow a more reliable determination of the background signal
from charmonium propagation in ordinary nuclear matter.

Even though the $\cpair$ pair is created with rather high momentum even at
threshold, it may be possible to observe reactions where the pair is captured
by the target nucleus, forming `nuclear-bound quarkonium' \cite{bts}. This
process should be enhanced in subthreshold reactions. There is no Pauli
blocking for charm quarks in nuclei, and it has been estimated there is a large
attractive van der waals potential binding the pair to the nucleus \cite{lms}.
The discovery of such qualitatively new bound states of matter would be a scoop
for any accelerator.

\section{Conclusions}

There are (at least) three central physics areas which require an
accelerator with the capabilities of ELFE as given in Table 1:
\begin{itemize}
\item[$\bullet$] The determination of hadron and nuclear wave functions.
\item[$\bullet$] Specifically nuclear effects: Color transparency \cite{ct},
cumulative phenomena \cite{bcdms} -- \cite{fermom}.
\item[$\bullet$] Charm(onium) production near threshold.
\end{itemize}

In addition to these core topics there are a number of areas where ELFE
can improve on presently available data, such as 
\begin{itemize}
\item[--] The nucleon structure function for $0.7\lsim x \lsim 1$,
\item[--] Higher twist corrections of the form $c(x)/Q^2$,
\item[--] $R=\sigma_L/\sigma_T$,
\item[--] The gluon structure function,
\item[--] Polarized structure functions.
\end{itemize}
Significant advances in these areas are, however, expected from other
experiments before ELFE starts operating.

Finally, we should keep in mind that the whole area of `confinement' physics
is very important but at present poorly understood in QCD. It includes open
questions like the influence of the QCD vacuum on scattering processes
\cite{nacht} and the foundations of the non-relativistic quark model (see,
\eg, \cite{diak,eps}). It is difficult to assess today what the progress will
be in this field. Nevertheless, it seems clear that systematic
measurements of non-perturbative wave functions as discussed above will form an
essential part of any serious effort to understand the hadron spectrum.   

{\bf Acknowledgements.} I am grateful to the organizers of this meeting for
their invitation to discuss physics in a stimulating atmosphere and a
delightful setting. I also wish to thank S. Brodsky and M. V\"anttinen, in
particular, for longstanding collaborations on the issues presented
above, as well as B. Kopeliovich and M. Strikman for several useful
discussions.

\end{document}